\documentclass[aip,jcp,reprint]{revtex4-1}
\usepackage{colordvi,epsfig,color,amsmath,amssymb}

\begin{document}

\def\be{\begin{equation}}
\def\ee{\end{equation}}
\def\bee{\begin{eqnarray}}
\def\eee{\end{eqnarray}}
\def\kb{k_{\rm B}}
\def\tilde{\widetilde}
\def\halb{\mbox{$\frac{1}{2}$}}
\def\with{\quad\mbox{with}\quad}
\def\und{\quad\mbox{and}\quad}
\newcommand{\bbbone}{{\mathchoice {\rm 1\mskip -4mu l}{\rm 1\mskip 
-4mu l}{\rm 1\mskip -4.5mu l}{\rm 1\mskip -5mu l}}}
\newcommand{\wn}{ cm$^{-1}$}
\newcommand{\ket}[1]{|#1\rangle}
\newcommand{\bra}[1]{\langle #1|}

\title{Hydration shell effects in the relaxation dynamics of photoexcited Fe-II
complexes in water}

\author{P.\ Nalbach$^{1,2}$, A.\ J.\ A.\ Achner$^{1,2}$, M.\ Frey$^{1,2}$, M.\ 
Grosser$^{1,2}$, W. Gawelda$^{2,3}$, A. Galler$^{2,3}$, T. Assefa$^{2,3}$, C.\
Bressler$^{2,3}$, and M.\ Thorwart$^{1,2}$}
\affiliation{$^1$I.\ Institut f\"ur Theoretische Physik,  Universit\"at
Hamburg, Jungiusstra{\ss}e 9, 20355 Hamburg, Germany \\
$^2$The Hamburg Centre for Ultrafast Imaging, Luruper Chaussee 149, 22761
Hamburg, Germany \\
$^3$European XFEL GmbH, Notkestra{\ss}e 85, 22607 Hamburg, Germany}

\date{\today}

\begin{abstract}
We study the relaxation dynamics of photoexcited Fe-II complexes dissolved in
water and identify the relaxation pathway which the molecular complex follows
in presence of a hydration shell of bound water at the interface between the
complex and the solvent. Starting from a low-spin state, the photoexcited
complex can reach the high-spin state via a cascade of different possible
transitions involving electronic as well as vibrational relaxation processes.
By numerically exact path integral calculations for the relaxational dynamics
of a continuous solvent model, we find that the vibrational life times of the
intermittent states are of the order of a few ps. Since the electronic
rearrangement in the complex occurs on the time scale of about 100 fs, we find
that the complex first rearranges itself in a high-spin and highly excited
vibrational state, before it relaxes its energy to the solvent via vibrational
relaxation transitions. By this, the relaxation pathway can be clearly
identified. We find that the life time of the vibrational states increases with
the size of the complex (within a spherical model), but decreases with the
thickness of the hydration shell, indicating that the hydration shell acts as an
additional source of fluctuations. 
\end{abstract}
\maketitle

\section{Introduction}

When a photoexcited molecule is placed in a polarizable solvent, it will relax 
its energy in presence of potentially strong interactions with its bath, i.e.,
its nearest neighbor solvent molecules. This interplay manifests itself already
in the properties of the steady state by the observed Stokes shift
between the absorption and emission energies of the solute, which typically 
reflect the rearrangement of the caging solvent around the excited solute
\cite{Fleming96,Ball08,Cramer99}. A pioneering femtosecond transient absorption
laser study of photoexcited NO in solid Ne and Ar rare gas matrices was capable
of extracting mechanistic movements of the caging rare gas atoms in combination
with model calculations \cite{Jeannin00,Vigliotti02}, but in liquid media this
connection to
the actual solvent movements in response to the creation of an excited state
dipole moment is inherently difficult to observe experimentally. Quantum
chemical calculations have meanwhile advanced and now permit simulating the
dynamic response inside a box containing the excited molecule itself and a
certain number of moving solvent molecules. In this way, simple
ions \cite{Pham11},  but also more complex molecules, such as aqueous
[Fe(bpy)$_3$]$^{2+}$  could be treated\cite{Daku10}. In a recent experiment,
Haldrup {\em et al.\/} have attempted to tackle this phenomenon exploiting
combined x-ray spectroscopies and scattering tools \cite{Haldrup12}. This
picosecond time-resolved experiment used x-ray absorption spectroscopy to
unravel the electronic changes visible around the Fe $K$ absorption edge. They 
occur concomitant to the geometric structural changes already extracted from the
extended x-ray absorption fine structure (EXAFS) region \cite{Gawelda09}. The
latter monitors the molecular changes around the central Fe atom. While
these
studies only shed light on the excited molecular dynamics within itself, the
recent study combined x-ray emission spectroscopy (XES) with x-ray diffuse
scattering (XDS) to obtain a picture of the internal electronic and structural
dynamics (via XES) simultaneously with the geometric structural changes in the
caging solvent shell. One surprising result from this experimental campaign
has yielded information about a density increase right after photoexcitation
(i.e., within the 100 ps time resolution of that study), which was fully in
line with the MD simulations of Ref.\ \onlinecite{Daku10}. They calculated a
change in the solvation shell between the low spin (LS) ground and high spin
(HS) excited state, which resulted in the expulsion of – on average – two water
molecules from the solvation shell into the bulk solvent. This showed up in the
XDS data as a density increase in the transient XDS pattern, and even the
quantitative analysis extracted an average density increase due to about two
water molecules expelled into the bulk solvent per photoexcited
[Fe(bpy)$_3$]$^{2+}$. 

This success has triggered the current theoretical study: If it is becoming
possible to experimentally gain new insight into guest-host interactions in
disordered systems like aqueous solutions, would it be possible to eventually
understand the influence of guest-host interactions on the dynamic processes
occurring within the solute? Indeed, aqueous [Fe(bpy)$_3$]$^{2+}$ is an ideal
model system for several reasons. Internally, it undergoes several ultrafast
transition processes involving correlations within the $3d$ orbitals: after
photoexcitation from the $^1$A$_1$ ground state into its singlet excited
metal-to-ligand charge transfer state ($^1$MLCT), it rapidly undergoes an 
intersystem crossing into the triplet manifold ($^3$MLCT) within about 30 fs
\cite{Cannizzo06}, and leaves the MLCT manifold in 120 fs \cite{Gawelda07}.
Femtosecond XAS studies observed the appearance of the finally accessed
HS $^5$T$_2$ state in less than 250 fs \cite{Bressler09,Lemke13}, which was also
confirmed by an ultrafast optical-UV transient absorption study
\cite{Consani09}. A very recent femtosecond XES study revealed the existence of
a metal-centered intermediate electronic state on the fly before the system
settles into the HS state \cite{Zhang14}. This electronic and spin-switching
process sequence starts from the LS ground state which is formed by six
paired electrons in the lower t$_{2g}$ level. Then, the cascade proceeds to
the HS excited state. There, the six electrons are distributed via
t$_{2g}^4$e$_g^2$ and both e$_g^2$ electrons with parallel spins to two of the
four t$_{2g}^4$ electrons. Overall, $S=2$ in the HS state (against
$S=0$ in the ground state) results. Such a transition is very common in Fe-II
based spin crossover (SCO) compounds, but little is understood about both the
internal dynamic processes involved as well as about the possible influence of
the solvent on this rapid spin-switching scheme. Indeed, the initially excited
MLCT manifold should
interact with the caging solvent molecules, but currently little is known about
the actual dynamic processes. This mystery motivates the calculations performed
in this work.

\begin{figure}[t!]
\centerline{\includegraphics[width=0.45\textwidth]{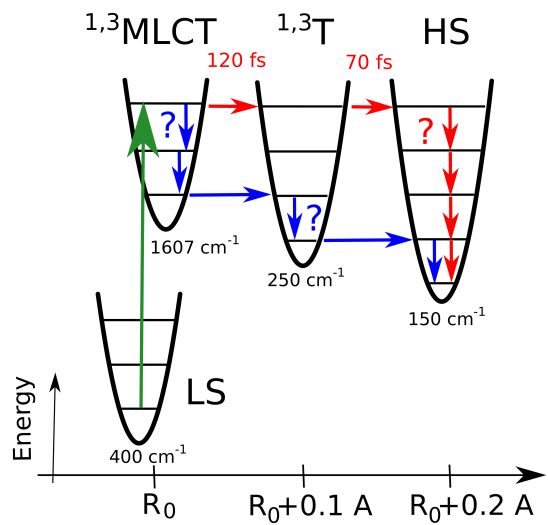}}
\caption{\label{fig1} Sketch of the energies of the LS, MLCT, T and the HS
state (not to scale). The details are given in the text. The two possible
relaxation pathways are indicated by the sequences of the red and blue arrows.
The unknown life time of the vibrational states is indicated by the blue
question mark and is determined in this work.}
\end{figure}

Here we investigate the energy relaxation dynamics in photoexcited aqueous
[Fe(bpy)$_3$]$^{2+}$ theoretically in order to provide a new view of the
short-time guest-host interactions in this complex sequence of relaxation. The
water molecules close to the compound are polarized and a hydration shell of
bound water is formed. On the one hand, this hydration shell may shield the
complex from polarization fluctuations
provided by the bulk water. On the other hand, it may also act as an additional
source of polarization fluctuations and thus enhance the relaxation process. 
To be specific here, we consider the case
of [Fe(bpy)$_3$]$^{2+}$ in water \cite{Bressler09,Haldrup12,Daku10}. The set of
states which are involved in the cascade of transitions from the LS to the HS
state is schematically shown in Fig.\ \ref{fig1}. Also, several intermediate
vibronic states of the complex are relevant \cite{Tuchagues04}.  An initial
photoexcitation (green solid arrow) brings the Fe-II
complex from the ground state of the LS configuration into an excited vibronic
state of a configuration of the metal-to-ligand-charge-transfer (MLCT) state.
The photoexcitation at $400$ nm provides an energy of about $3.1$ eV or
$\sim 25000$ \wn. More precisely, a state on the $^1$MLCT manifold is initially
excited, but rapidly undergoes an intersystem crossing into the triplet manifold
($^3$MLCT) within about 30 fs \cite{Gawelda07}. The two manifolds are similar in
their vibrational frequencies and correspond to the skeleton mode of bpy in the
MLCT configuration. This mode has a rather high vibrational frequency of
$\Omega_{\rm MLCT}=1607$ \wn and its vibrational ground state has an energy of
about $18000$ \wn. 
Hence, the photoexcitation populates mostly the vibrational state $\ket{4}_{\rm
MLCT}$ with a quantum number $\nu_{MLCT}=4$. 

The relaxation out of this state can now occur via two alternative relaxation
pathways. Elements of these pathways are known, but the path which is
eventually chosen by the system is not fully understood in detail up to present.
On the one hand, the relaxation can proceed via energetically lower-lying
vibrational states on the MLCT manifold, i.e., following $\ket{4}_{\rm
MLCT}\to\ket{3}_{\rm MLCT} \to \dots$ (the blue path, see the sequence of blue
arrows in Fig.\ \ref{fig1}). In fact, the available MLCT states form a broad
manifold of metal-centered states \cite{Bressler09}. From the MLCT ground state,
the energy could be transferred to a vibrationally excited state of one of 
the metal-centered triplet states ($^{1/3}$T). In the T state, the Fe-N bond
length increases, such that the Fe-II complex expands by about $0.1$ \AA. This
molecular configuration has a vibrational energy gap \cite{Tuchagues04} of
$\Omega_{\rm T} \sim 250$ \wn which corresponds to a vibrational mode of the
Fe-N bond. 
It is experimentally well-established that the transfer from the MLCT manifold
to the intermediate T states 
occurs in about $120$ fs. The system would reach the
vibrational ground state of the T configuration via a sequence of vibrational
relaxation steps. From the T-vibrational ground state, the energy would be
transferred to a vibrational excited state of the HS configuration.  Its HS
vibrational ground state has an energy of $\sim 4000$ \wn. The vibrational
energy gap is again determined by a vibrational mode of the
Fe-N bond and is estimated \cite{Tuchagues04} as $\Omega_{\rm HS} \sim 150$
\wn. 
It is established that the transfer from the T to the HS state
occurs within $70$ fs \cite{Bressler09,Zhang14}. Along with this occurs another
rearrangement of the compound which results in an effective growth of the
molecule (and thus somewhat the caging cluster) of $0.2$ \AA. 
After vibrational relaxation in the HS state, the system would reach its HS
ground state configuration within $960$ fs \cite{Gawelda07}. 

The second possible pathway (the red path, see the sequence of red arrows in
Fig.\ \ref{fig1}) would start in a highly excited vibrational state on the MLCT
manifold as before. Without performing a vibrational relaxation transition
within the MLCT manifold, it directly yields to a highly excited vibrational
state on the T manifold within $120$ fs and continues again without a
vibrational relaxation transition to another highly excited vibrational state of
the HS configuration within $70$ fs. From there, the complex
relaxes into the HS ground state via vibrational transitions and removal of the
corresponding energy into the hydration shell and bulk water within $960$ fs
\cite{Gawelda07}.

The final HS to LS relaxation occurs in $665$ ps \cite{Gawelda07}.  

Both scenarios would allow the system to reach the HS electronic manifold 
within roughly $200$ fs via several intermediate states. The initial energy is
intermittently stored within molecular vibrations but finally transferred out of
the complex into the solvent environment.  The Fe-II complex expands, since 
the Fe-N bond lengths increase, when the compound is 
excited from the LS to HS state. Here, we assume that Fe-N stretching and
bending modes are dominant. 

What is unknown from the experimental perspective, is the vibrational life
times of the intermediate vibrational electronic states (blue and red question
marks in Fig.\ \ref{fig1}). For instance, if the highly excited vibrational
states on the MLCT manifold live long enough such
that the transfer to the T-manifold can occur within $120$ fs, the system would
most likely choose the red pathway. On the other hand, if the highly excited
vibrational states on the MLCT manifold rapidly relax within $120$ fs to the
MLCT ground state, the system would prefer to follow the blue relaxation
pathway. 

To decide this question from a theoretical point of view, we follow a simplified
model description which is accurate enough such that a clear qualitative answer
follows. For this, we establish a model
of a quantum mechanical two-state system which describes a bath-induced
vibrational relaxation from an excited vibrational state to the ground
state on a generic manifold. We thereby
model the environmental polarization fluctuations including the effects of a
hydration shell in terms of a refined Onsager model combined with a Debye
relaxation picture \cite{Abe}. A crucial aspect here is that we include the
bulk solvent {\em and\/} the hydration shell on the same footing in terms of a
continuum description of environmental Gaussian modes. This model allows us
easily to modify the radius of the solvated complex (taken as a sphere in this
work) and the thickness of the surrounding hydration shell. Within this
simplified model, we determine the energy relaxation rate for several
representative vibrational modes including the Fe-N stretching and bending modes
in dependence of the Fe-N bond length and the
hydration shell thickness. Technically, we use numerically exact real-time path
integral simulations on the basis of a fluctuational spectrum which is highly
structured and far from being Ohmic. Such a ``slow'' bath reflects the
similar physical time scales on which the vibrational relaxation transitions
within a vibrational manifold and the polarization fluctuations of the
surrounding water occur. The highly non-Ohmic form (see below) of the bath
spectral densities a priori calls for the use of an advanced theoretical method
beyond the standard Markov-approximated dynamical Redfield equations. 
\begin{figure}[t!]
\centerline{\includegraphics[width=0.45\textwidth]{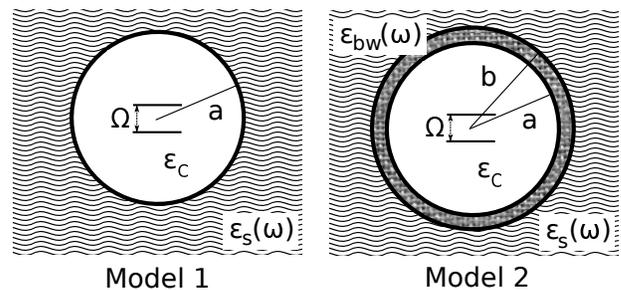}}
\caption{\label{fig2} Sketch of continuum dielectric models for the
complex-bound-water-solvent system, see text for details. $a$ denotes the
radius of the inner sphere, while $b$ refers to the radius of the outer sphere.
$\varepsilon_{\rm s}(\omega)$ is the frequency-dependent complex dielectric
function of the continuum bulk water modes. $\varepsilon_{\rm bw}(\omega)$ is
the frequency-dependent complex dielectric function associated with the bound
water shell. $\varepsilon_c=1$ is the dielectric constant of the vacuum inside
the
cavity.}
\end{figure}

We find vibrational energy relaxation times on generic manifolds in the
range of $2 -8$ ps depending on the Fe-N bond lengths and the hydration shell
thickness. For this, we tune the vibrational frequencies which are determined
by the curvature of the manifolds over a relevant parameter range. 
We can determine the modes with fastest energy relaxation which dominates the
energy relaxation dynamics of the Fe-II complex since internal energy
redistribution is likely much faster. Most importantly, we observe that the
vibrational relaxation times within a manifold are much longer than the typical
time scales of a few hundred fs during which the HS state is formed. Two effects
are competing here. A complex with a smaller radius of the solvation sphere
brings
the environmental fluctuations spatially closer to the complex and thus results
in a faster decay. However, in turn, the stronger Fe-N bond results in
larger mode frequencies. Overall, the calculated life times of the
vibrationally excited states in the ps regime clearly show that the
vibrational life times are much longer than the complex overall needs to reach
the HS state, which are less than $200$ fs. Thus, we can conclude that the
energy relaxation basically occurs via the ``red pathway'' after the
complex has reached the HS state and vibrationally relaxes into the ground
state. 

\section{Model}

To determine the life time of the excited vibrational states, we formulate a
minimal model in form of a quantum two-level system which is immersed in its
solvent environment (model 1) and is, in addition, surrounded by a hydration
shell (model 2). After expansion of the Fe-II complex, the stretching and
bending modes \cite{Sousa2013} involving the Fe-N bond change their respective
vibrational frequency. We investigate their relaxation dynamics independently
and use the spin-boson Hamiltonian \cite{Weiss} as a minimal model, i.e., 
\begin{equation} \label{totham}
 H=\frac{\hbar \Omega}{2} \sigma_z + \hbar \sigma_x \sum_j c_j (b_j
+b_j^\dagger) + \sum_j \hbar \omega_j b_j^\dagger b_j \, .
\end{equation}
Here, the Pauli matrix $\sigma_z$ contains the ground state $\ket{g}$ and the
excited state $\ket{e}$ between which we investigate the relaxation
transitions. The two states are separated in energy by the vibrational
frequency $\Omega$. The bath modes produce Gaussian fluctuations stemming from
harmonic oscillators with frequencies $\omega_j$, the corresponding creation
and destruction operators of the bath modes are denoted as $b_j^\dagger$ and
$b_j$. The fluctuations induce transitions in the system via the Pauli matrix
$\sigma_x$. They can be characterized by a single function \cite{Weiss}, the
spectral density
\begin{equation}
 J(\omega)=2 \pi \sum_j c_j^2 \delta(\omega-\omega_j) \, .
\end{equation}
It provides the spectral weight contained in the fluctuations at frequency
$\omega$ which are provided by a Gaussian bath at thermal equilibrium at a
given fixed temperature $T=1/(k_B \beta)$. The correlation function of the
quantum bath fluctuations $\xi(t)$ is given by ($t>0$)
\begin{equation}
 \langle \xi(t) \xi (0) \rangle = \frac{1}{\pi} 
\int_0^\infty d\omega J(\omega) \left[
\coth \frac{\hbar \omega \beta}{2} \cos \omega t - i \sin \omega t 
\right] \, .
\end{equation}
This quantity determines the relaxation and dephasing rates \cite{Weiss}. 
In this work, we consider several representative Fe-N stretching and bending
modes with the frequencies $\Omega=$ 60, 120, 150, and 250 cm$^{-1}$. 
Moreover, we use a continuum description of the solvent (bulk) 
water and the hydration shell following Gilmore and McKenzie \cite{Gilmore08}.
The key quantity to characterize the environment, i.e., the spectral
distribution $J(\omega)$ of the fluctuations, is determined in terms of the
standard Onsager model of polarization fluctuations of the solvent water
molecules. Their relaxation properties are described within a Debye relaxation
picture \cite{Abe}. In this approach, the spectral density is related to the
continuum dielectric function $\varepsilon (\omega)$ of the host material. 

To be more specific, we consider two different situations \cite{Gilmore08}, see
Fig.\ \ref{fig2}: 
In model 1, we assume that the complex with its vibrational mode is placed
inside a vacuum spherical cavity of radius $r_a$ with a dielectric constant
$\varepsilon_c=1$. This is situated in a continuum of bulk water modes with a
frequency-dependent complex dielectric function $\varepsilon_{\rm s}(\omega)$.
In model 2, we add to model 1 an outer sphere with
radius $r_b>r_a$. The shell formed by the two spheres describes the bound
water or hydration shell in terms of a second frequency-dependent
complex dielectric function $\varepsilon_{\rm bw}(\omega)$. This model allows
us to determine the relaxation rates also for varying the radii $r_a$ and
$r_b$ independently. Throughout this work, we set $T=300$ K. 

\subsection{Model 1: Bulk water} 
Following Gilmore and McKenzie \cite{Gilmore08,Gilmore05}, one can calculate
the reaction field by solving Maxwell's equation for the particular geometry
shown in Fig.\ \ref{fig2}. This yields the spectral
density
\begin{eqnarray}
 J_1(\omega) &=& \frac{(\Delta \mu)^2}{2\pi \varepsilon_0 r_a^3} {\rm Im } 
\frac{\varepsilon_{\rm s}(\omega)-1}{2\varepsilon_{\rm s}(\omega)+1} \nonumber
\\
& = & \frac{(\Delta \mu)^2}{2\pi \varepsilon_0 r_a^3} \frac{6(\varepsilon_{{\rm
s},0}-\varepsilon_{{\rm s},\infty})}{(2\varepsilon_{{\rm
s},0}+1) (2\varepsilon_{{\rm s},\infty}+1)} \frac{\omega \tau_{\rm s}}{\omega^2
\tau_{\rm s}^2+1} \, ,
\end{eqnarray}
with the respective transition dipole moment $\Delta \mu$ of the vibration,
$\varepsilon_{{\rm s},0}$ being the static dielectric constant of the
bulk solvent, $\varepsilon_{{\rm s},\infty}$ being the high-frequency dielectric
constant of the bulk solvent, and
\begin{equation}
 \tau_{\rm s} = \frac{2\varepsilon_{{\rm s},\infty}+1}{2\varepsilon_{{\rm
s},0}+1} \tau_{\rm D,s}
\end{equation}
and $\tau_{\rm D,s}$ is the Debye relaxation time of the solvent. For water, we
have $\varepsilon_{{\rm s},0}=78.3, \varepsilon_{{\rm s},\infty}=4.2$ and 
$\tau_{\rm D,s}=8.2$ ps. 

Here, we are interested in the dependence of the spectral density on the cavity
volume determined by its radius $a$ and we thus collect all constants in a
 prefactor. We arrive at 
\begin{equation}
J_1(\omega) = \frac{\alpha_1}{a^3} \frac{\omega}{\omega^2 \tau_{\rm s}^2+1} \, ,
\end{equation}
with
\begin{equation}
 \alpha_1 = \frac{1}{2\pi \hbar}\frac{(\Delta \mu)^2}{2\pi \varepsilon_0 a_0^3}
\frac{6(\varepsilon_{{\rm
s},0}-\varepsilon_{{\rm s},\infty})}{(2\varepsilon_{{\rm
s},0}+1) (2\varepsilon_{{\rm s},\infty}+1)}\tau_{\rm s}\, ,
\end{equation}
where $a_0$ is the typical length scale of the problem and where the now
dimensionless radius $a=r_a/a_0$ is measured in units of $a_0$. We fix this to
$a_0=1$ \r{A} throughout this work. The spectrum is purely Ohmic \cite{Weiss}
with a cut-off frequency given by 
$\omega_{\rm c,s}= 1/\tau_{\rm s}$. For our considerations, we fix the dipole
moment to a typical value of $\Delta \mu = 1 $ D $= 3 \times 10^{-30}$ Cm.
Collecting all parameters yields $\alpha_1\approx 5$ for bulk water.
\begin{figure}[t]
\includegraphics[width=85mm,angle=0]{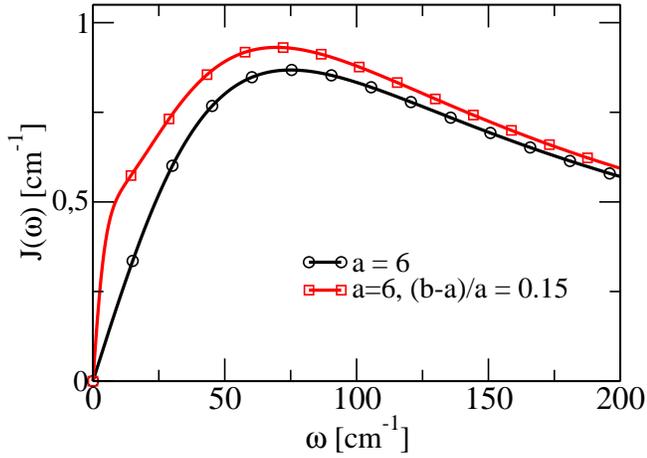}
\caption{\label{fig3}Spectral densities for model 1 (black line, circles) and
 model 2 (red line, squares) for water and for a cavity radius of $r_a=6$ \AA\/
and a
relative shell thickness of $(b-a)/a=0.15$.}
\end{figure}
Fig. \ref{fig3} shows $J_1(\omega)$ of model 1 for the case $a=6$
(corresponding to $r_a=6$ \AA). Maximal spectral weight is observed at roughly
70 cm$^{-1}$. Hence, it is clear that the resulting bath correlation times are
comparable to or exceed internal system 
periods. This also prevents us from using a standard Markov approximation a
priori, since a correlated and non-Markovian dynamics can in principle be
expected \cite{Nalbach-JCP-2010} (see below).

\subsection{Model 2: Bulk water plus hydration shell} 
We also include the hydration shell of bound water and do this by a second
sphere with outer radius $r_b=b a_0$ with $b$ being the corresponding
dimensionless number. We assume that the hydration shell is thin
relative to the radius of the inner sphere and may then perform a Taylor
expansion in the relative shell thickness $(b-a)/a$. The resulting spectral
density \cite{Gilmore08} is
\begin{equation}
  J_2(\omega)=J_1(\omega) + J_{\rm bw}(\omega)\, , 
\end{equation}
with
\begin{eqnarray}
 J_{\rm bw} (\omega) &=& \frac{(\Delta \mu)^2}{2\pi \varepsilon_0 r_a^3}
\frac{1}{(2\varepsilon_{\rm s}(\omega)+1)^2}\left(
1+\frac{2\varepsilon_{{\rm s},0}}{|\varepsilon_{\rm bw}(\omega)|^2}\right) 
{\rm Im } \varepsilon_{\rm bw}(\omega) \, ,\nonumber \\
\end{eqnarray}
where $\varepsilon_{\rm bw}(\omega)$ is the complex dielectric function of the
bound water layer. Within the Debye relaxation model, we find
\begin{equation}
 J_{\rm bw} (\omega) = \frac{\alpha_{\rm bw}}{a^3} \frac{b-a}{a} 
\frac{\omega}{{\omega}^2 \tau_{\rm bw}^2+1}
\end{equation}
with
\begin{equation}
 \alpha_{\rm bw} = \frac{1}{2\pi \hbar}\frac{3(\Delta \mu)^2}{2\pi \varepsilon_0
a_0^3}
\frac{(\varepsilon_{{\rm
bw},0}^2+\varepsilon_{{\rm s},0}^2) (\varepsilon_{{\rm
bw},0}-\varepsilon_{{\rm bw},\infty})}{\varepsilon_{{\rm
bw},0}^2 (2\varepsilon_{{\rm s},0}+1)^2} \tau_{\rm bw}\, .
\end{equation}
Here, we have the static dielectric constant $\varepsilon_{{\rm bw},0}$ and the
high-frequency dielectric constant $\varepsilon_{{\rm bw},\infty}$ of the bound
water layer. 
From generic considerations \cite{Gilmore08}, one may infer that the relaxation
time of the
bound water shell is one order of magnitude large than the solvent 
relaxation time, i.e., we set $\tau_{\rm bw}=10\tau_{\rm s}$. 
Likewise, we know \cite{Gilmore08} that
$\varepsilon_{{\rm s},0}\gg \varepsilon_{{\rm s},\infty}$.
Moreover, $\varepsilon_{{\rm
bw},0} \gg \varepsilon_{{\rm bw},\infty}$ and $\varepsilon_{{\rm s},0}\gg
\varepsilon_{{\rm bw},0}$. Hence, we may use this and set $\varepsilon_{{\rm
bw},0}=1$ to obtain
\begin{equation}
 \alpha_{\rm bw} = \frac{1}{2\pi \hbar} \frac{3(\Delta \mu)^2}{2\pi
\varepsilon_0 a_0^3} \frac{1}{4} \tau_{\rm bw} \, .
\end{equation}
For the parameters mentioned, we find $\alpha_{\rm bw}\approx 118$. 
\begin{figure}[t]
\centerline{\includegraphics[width=0.48\textwidth]{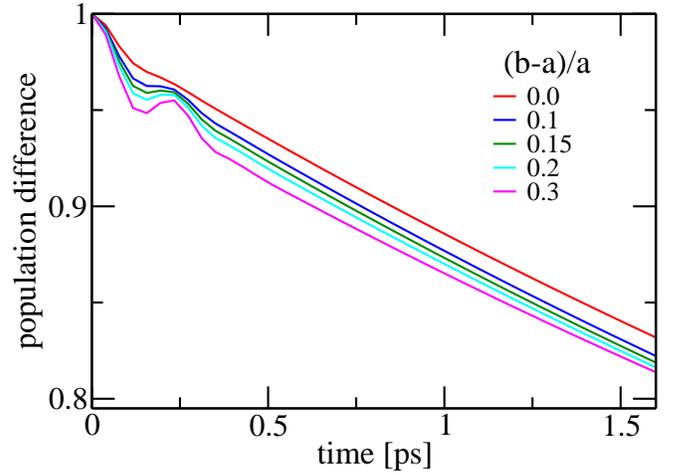}}
\caption{\label{fig3a}Time evolution of the population difference $P(t)$ for
$\Omega=150$ \wn for $T=300$ K for different values $(b-a)/a$ of the shell
thickness.}
\end{figure}

Fig. \ref{fig3} shows $J_2(\omega)$ for these parameters and for $a=6$ and
$(b-a)/a=0.15$. Again, maximal spectral weight is observed at
roughly 70 cm$^{-1}$. In general, the spectral weight of model 2 is higher than
of model 1. This already indicates that within the continuum approach, the
bound water shell acts as an additional source of fluctuations and not as a
spectral filter for the continuous bulk modes. Hence, the calculated relaxation
times for model 2 will be faster than for model 1.

Moreover, it is clear that the vibrational life times on the MLCT manifold are
much larger since there the spectral weight of the solvent environmental modes
 around the frequency of $\Omega_{\rm MLCT}=1607$ \wn\  is strongly suppressed
(in fact, we do not consider the vibrational relaxation around this frequency
in this work).  

\section{Real-time dynamics of the relaxation transitions}
%
\begin{figure}[t]
\centerline{\includegraphics[width=0.48\textwidth]{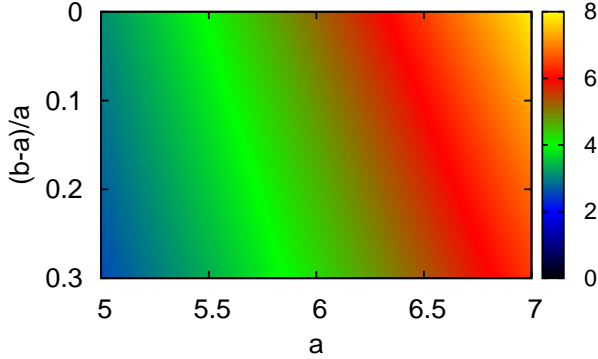}}
\caption{\label{fig4} Relaxation time (color scale in ps) of the excited
vibrational state for varying radius $r_a=a a_0$ with $a_0=1 $ {\AA}
\hspace{1ex} and for
varying relative shell thickness $(b-a)/a$ for $\Omega=$150 cm$^{-1}$. Model 1
with no hydration shell is contained via the cut along the line $(b-a)/a=0$.}
\end{figure}
%
To investigate the quantum relaxation dynamics of the two
vibrational states under the influence of environmental
fluctuations, we employ the numerically exact
quasi-adiabatic propagator path-integral (QUAPI) \cite{Makri-JCP-1995} scheme
which we have
extended to allow treatment of multiple baths \cite{Nalbach-NJP-2010}.
Specifically, QUAPI is able to treat highly structured and non-Markovian baths
efficiently \cite{Nalbach-JPB-2012,MujicaMoQPRL13,MujicaPRE2013}. It determines
the time dependent statistical operator $\rho (t)$ which is obtained after the
harmonic bath modes have been integrated over. We briefly summarize here the
main ideas of this well-established method and refer to the literature for
further details. The algorithm is 
based on a symmetric Trotter splitting of the short-time propagator ${\cal
K}(t_{k + 1}, t_k)$ for the full Hamiltonian Eq.\ (\ref{totham}) into a part
depending on the system 
Hamiltonian alone and a part involving the bath and the coupling term. The
short-time 
propagator gives the time evolution over a Trotter time slice $\delta t$. This
splitting in discrete time steps is exact in the limit $\delta t \to 0$, i.e.,
when the discrete time evolution approaches the limit of a continuous
evolution. For any
finite time slicing, it introduces a finite Trotter error which has to be
eliminated by choosing $\delta t$ small enough such that convergence is
achieved. On the other side, the environmental degrees of freedom generate
correlations being non-local in time. We want to avoid any Markovian
approximation at this point and take these correlations into account on an
exact footing. We may, however, use the fact that for any finite temperature,
these
correlations decay exponentially quickly on a time scale denoted as the memory
time scale. 
The QUAPI scheme now defines an object called the reduced density tensor. It
corresponds to an extended quantum statistical operator of the system which is
nonlocal in time since it lives on this memory time window. By this, one can
establish an iteration scheme by disentangling the dynamics in order to extract
the time evolution of this object. All correlations are fully included over
the finite memory time  $\tau_{\rm mem} = K \delta t$, but are neglected for
times beyond $\tau_{\rm mem}$. To obtain numerically exact results, we have to
increase accordingly the memory parameter $K$ until convergence is found. The
two strategies to achieve convergence, i.e., minimize $\delta t$ but maximize
$\tau_{\rm mem} = K \delta t$, are naturally counter-current,
 but nevertheless convergent results can be obtained in a wide range of
parameters, including the cases presented in this work.

\section{Results}
\begin{figure}[t]
\centerline{\includegraphics[width=0.48\textwidth]{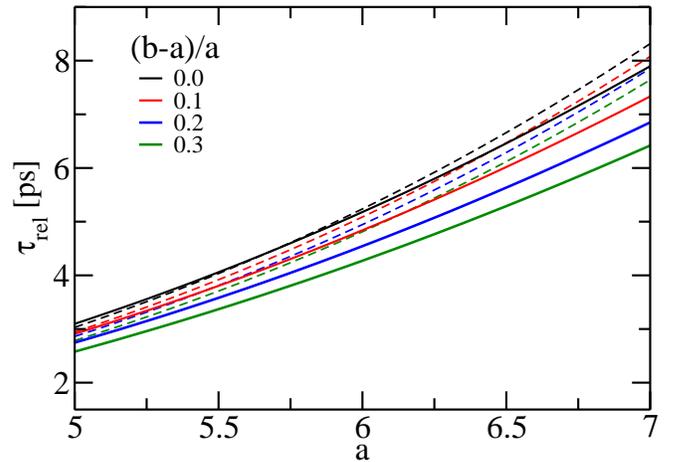}}
\caption{\label{fig5} Solid lines: Cut through the 2D plot of Fig.\ \ref{fig4}
along the lines $(b-a)/a=0, 0.1, 0.2$ and $0.3$. The dashed lines indicate the
results of the vibrational life times calculated within a Born-Markov
approximation (see text).}
\end{figure}
At first, we consider modes with a vibrational frequency of $\Omega=$150
cm$^{-1}$. We determine the difference $P(t)=\langle \sigma_z\rangle_t = {\rm
tr} [\rho(t) \sigma_z]$ of the populations of the ground and the excited states.
We start out from the initial preparation of the excited state, i.e.,
$\rho(0)=\ket{e}\bra{e}$. Fig.\ \ref{fig3a} shows
examples of the relaxation dynamics for the environmental models 1 and 2 for
different values of the shell thickness $(b-a)/a$. We
mainly observe exponential relaxation on a time scale of a few picoseconds.
For an increasing shell thickness, a tendency towards a decaying oscillatory
dynamics appears. A pronounced oscillation with a period of $\sim 250$ fs
develops for the largest thickness considered. 

To quantify the decay in terms of life times of the excited state, we extract
from the time evolution the corresponding rate by a fit to an exponential. 
Fig.\ \ref{fig4} shows the relaxation time in ps (colour scale) as a function of
the radius $a$ of the complex varying it between 5 to 7 \AA\ and the relative
shell thickness $(b-a)/a$ varying it between 0 to 30\%, which is consistent
with the numerical findings of Ref.\ \onlinecite{Daku10}.  
The plot shows results of both, models 1 and 2 (model 1 corresponds to the line
with $(b-a)/a=0$). The data for $(b-a)/a=0, 0.1, 0.2$ and $0.3$ are shown again
in Fig.\ \ref{fig5} for better readability. The calculated relaxation times or
life times of the excited state vary from $2$ to $8$ ps. For a larger complex
radius, the 
life time increases as expected since the prefactor of the spectral
density decreases proportional to $1/r_a^3$. This reflects the assumption that
the effective transition 
dipole sits in the center of the sphere and an increasing complex pushes
the solvent fluctuations further away. This reduces their strength due to the
distance dependence of the dipolar coupling. Moreover, the life 
times decrease with increasing hydration shell thickness. Thus, the hydration
shell does not act as a shield from bulk solvent fluctuations but acts as an 
additional source of fluctuations instead.

Fig.\ \ref{fig5} also shows the results of the vibrational life times
calculated within a Born-Markov approximation \cite{Weiss}. The inverse life
time or the relaxation rate can be obtained after expanding the transition
rates in a master equation approach up to lowest order in the system-bath
interaction, together with a Markovian approximation of the bath-induced 
correlations. This corresponds effectively to only including single-phonon
transitions in the bath. The inverse vibrational life time then follows as 
\begin{equation}
 \tau^{-1} = J(\Omega) \coth \frac{\hbar \Omega \beta}{2} \, .
\end{equation}
As is shown by the dashed lines in Fig.\ \ref{fig5}, significant deviations
from the exact life times occur and the approximated life times are
overestimated by up to $10\%$. 

\begin{figure}[t]
\centerline{
\includegraphics[width=0.49\textwidth]{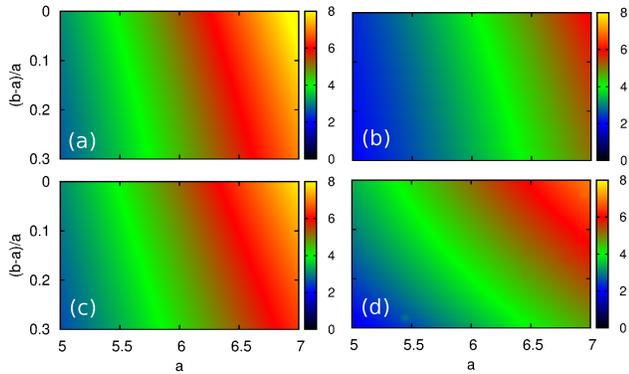}}
\caption{\label{fig6} Vibrational life times (color scale in ps) for
$\Omega=60$ \wn (a), $120$ \wn (b), $150$ \wn (c) and $250$ \wn (d) for models
1 and 2 for $T=300$ K.}
\end{figure}

Next, we show the results for the calculated life times for other vibrational
frequencies, i.e., for $\Omega=60, 120, 150$, and $250$\wn\ in Fig. \ref{fig6}.
These values span the regime of the vibrational frequencies for the Fe-N
stretching and bending modes in the LS and HS state \cite{Sousa2013}. Note that
the frequencies are comparable to or larger than the frequency for which the
maximal spectral weight in the environmental fluctuation spectrum occurs.
Hence, the energy relaxation dynamics occurs in the regime in which 
non-Markovian multi-phonon transitions already are noticeable 
\cite{Nalbach-JCP-2010}. We note that for larger values of $\Omega$, no
convergent results have been achieved, which is a further strong indication of
non-Markovian behavior. 

\section{Conclusions}

We observe that under the assumption of equal strengths of the coupling to the 
environmental fluctuations, all Fe-N stretching and bending modes in the LS and
HS state exhibit quite similar vibrational life times on the order of 5
ps. The vibrational energy gap has been modified from $60$ to $250$ \wn and all
cases show similar results. An increased radius of the complex results in a
larger life time since the fluctuating solvent molecules are moved further
outside. A finite hydration shell thickness reduces the vibrational life times
noticeably. 

Our results indicate that all vibrational modes contribute similarly to
the energy relaxation after initial photoexcitation. At the
same time, all vibrational modes live too long in order to relax the energy
already in the MLCT or the T state 
(assuming here the vibrational modes being identical to the modes in
the LS state). Hence, the energy after the photoexcitation is first rapidly
transfered from a highly excited vibrational MLCT state to a highly excited
vibrational T state and then further to a highly excited vibrational HS state
within about less than $200$ fs. Only then, the full excess energy is
dissipated while the electronic subsystem is in the HS state. Hence, the system
follows the ``red relaxation pathway'' sketched in Fig.\ \ref{fig1}. 

Energy redistribution within more molecular vibrational states is not included
in our simplified model. Assuming the excess energy initially equally
distributed among
the Fe-N stretching and bending modes \cite{Sousa2013}, each mode gets roughly
an excitation energy of 440 cm$^{-1}$. This implies that roughly two
excitations of the mode $\Omega=$ 250 cm$^{-1}$ and
up to three or four excitations of the mode with $\Omega=$ 120 cm$^{-1}$ and
$\Omega=$ 150 cm$^{-1}$ occur. Thus, the total equilibration time of the complex
after photoexcitation roughly follows as three times $5$ ps which yields a
value of $15$ ps. These results could be experimentally verified by ultrafast
spectroscopy of the intermediate MLCT and T states.

\section{Acknowledgments}

We acknowledge financial support by the DFG Sonderforschungsbereich 925
``Light-induced dynamics and control of correlated quantum systems'' (projects
A4 and C8) and by the DFG excellence cluster ``The Hamburg
Center for Ultrafast Imaging''. CB, AG, WG acknowledge funding by the
European XFEL.


\begin{thebibliography}{99}
\bibitem{Fleming96}G.R. Fleming and M. Cho, Annu. Rev. Phys. Chem. {\bf 47},
109 
(1996).

\bibitem{Ball08}P. Ball, Chem. Rev. {\bf 108}, 74 
(2008).

\bibitem{Cramer99}C. Cramer and D. Truhlar, Chem. Rev. {\bf 99}, 2161 
(1999).

\bibitem{Jeannin00}C. Jeannin, M.T. Portella-Oberli, S. Jiminez, F. Vigliotti,
B. Lang, and M. Chergui, Chem. Phys. Lett. {\bf 316}, 51 
(2000). 

\bibitem{Vigliotti02} F. Vigliotti, L. Bonacina, M. Chergui, G. Rojas-Lorenzo,
and J. Rubayo-Soneira, Chem. Phys. Lett. {\bf 362}, 31 
(2002).

\bibitem{Pham11}
V.-T. Pham, Thomas J. Penfold, R. M. van der Veen, F. Lima, A. El Nahhas, S.
Johnson, R. Abela, C. Bressler, I. Tavernelli, C. J. Milne, and M. Chergui,
J. Am. Chem. Soc. {\bf 133}, 12740 
(2011).

\bibitem{Daku10} L.M.L. Daku and A. Hauser, J. Phys. Chem. Lett. {\bf 1},
1830 
(2010).

\bibitem{Haldrup12}
K. Haldrup, G. Vank\'{o}, W. Gawelda, A. Galler, G. Doumy, A. M. March, E. P.
Kanter, A. Bordage, A. Dohn, T. B. van Driel, K. S. Kjaer, H. T. Lemke, S. E.
Canton, J. Uhlig, V. Sundstr\"om, L. Young, S. Southworth, M. M. Nielsen, and C.
Bressler, J. Chem. Phys. {\bf A116}, 9878 
(2012). 

\bibitem{Gawelda09} W. Gawelda, V.-T. Pham, R. M. van der Veen, D. Grolimund, R.
Abela, M. Chergui, and C. Bressler, J. Chem. Phys. {\bf 130}, 124520 (2009).

\bibitem{Cannizzo06}A. Cannizzo, F. Van Mourik, W. Gawelda, M. Johnson, F. M. F.
deGroot, R. Abela, C. Bressler, and M. Chergui, Angew. Chem. Intl. Ed. {\bf 45},
3174 
(2006).

\bibitem{Gawelda07} W. Gawelda, A. Cannizzo, V.-T. Pham, F. Van Mourik, C.
Bressler, and M. Chergui, J. Am. Chem. Soc. {\bf 129}, 8199 
(2007).

\bibitem{Lemke13} H. T. Lemke, C. Bressler, L.X. Chen, D.M. Fritz, K.J.
Gaffney, A. Galler, W. Gawelda, K. Haldrup, R. W. Hartsock, H. Ihee, J. Kim, K.
H. Kim, J. H. Lee, M. M. Nielsen, A. B. Stickrath, W. Zhang, D. Zhu, and M.
Cammarata, J. Phys. Chem. {\bf A117}, 735 
(2013). 

\bibitem{Bressler09} Ch. Bressler, C. Milne, V.-T. Pham, A. ElNahhas, R.M. van
der Veen, W. Gawelda, S. Johnson, P. Beaud, D. Grolimund, M. Kaiser, C.N.
Borca, G. Ingold, R. Abela, and M. Chergui, Science {\bf 323}, 489 (2009).

\bibitem{Consani09} C. Consani, M. Pr\'{e}mont-Schwarz, A. El Nahhas, C.
Bressler, F. Van Mourik, and M. Chergui, Angew. Chem. Int. Ed. {\bf 48},
7184 
(2009). 

\bibitem{Zhang14} W Zhang, R. Alonso-Mori, U. Bergmann, C. Bressler, M. Chollet,
A. Galler, W. Gawelda, R. G. Hadt, R. W. Hartsock, T. Kroll, K. S. Kj{\ae}r, K.
Kubi\v{c}ek, H. T. Lemke, H. W. Liang, D. A. Meyer, M. M. Nielsen, C.
Purser, J. S. Robinson, E. I. Solomon, Z. Sun, D. Sokaras, T. B. van Driel, G.
Vank\'{o}, T. Weng, D. Zhu, and K. J. Gaffney, Nature (in press 2014).

\bibitem{Tuchagues04}J.P. Tuchagues, A. Bousseksou, G. Molnar, J.J. McGarvey,
and F. Varret, Top. Curr. Chem. {\bf 235}, 55 (2004).

\bibitem{Abe}A. Nitzan, {\em Chemical Dynamics in Condensed Phases: Relaxation,
Transfer, and Reactions in Condensed Molecular Systems} (Oxford University
Press, Oxford, 2006).

\bibitem{Sousa2013} C. Sousa, C. de Graaf, A. Rudaskyi, R. Broer, J. Tatchen. M.
Etinski, and C. M. Marian, Chem. Eur. J. {\bf 19}, 17541 (2013). 

\bibitem{Weiss}U. Weiss, {\em Quantum Dissipative Systems\/}, 4th ed.\
(World Scientific, Singapore, 2012).

\bibitem{Gilmore08}J. Gilmore and R.H. McKenzie, J.\ Phys. Chem. A {\bf 112},
2162 (2008). 

\bibitem{Gilmore05}J. Gilmore and R.H. McKenzie, J.\ Phys.: Condens. Matter {\bf
17}, 1735 (2005). 

\bibitem{Nalbach-JCP-2010} P. Nalbach and M. Thorwart, J. Chem. Phys. {\bf 132},
194111 (2010).

\bibitem{Makri-JCP-1995}  N. Makri and D. E. Makarov, J. Chem. Phys. {\bf 102},
4600 (1995); ibid., 4611 (1995).

\bibitem{Nalbach-NJP-2010} P. Nalbach, J. Eckel, and M. Thorwart, New J. Phys.
{\bf 12}, 065043 (2010).

\bibitem{MujicaMoQPRL13} C.A. Mujica-Martinez, P. Nalbach, and M. Thorwart,
Phys. Rev. Lett. {\bf 111}, 016802 (2013).

\bibitem{Nalbach-JPB-2012} P. Nalbach and M. Thorwart, J. Phys. B: At. Mol. Opt.
Phys, {\bf 45}, 154009 (2012).

\bibitem{MujicaPRE2013} C. A. Mujica-Martinez, P. Nalbach, and M. Thorwart,
Phys. Rev. E {\bf 88}, 062719 (2013).

\end{thebibliography}
\end{document}